\documentclass[11pt,a4paper]{article}
\usepackage[utf8]{inputenc}
\usepackage{authblk}
\usepackage{amsmath,amsfonts,amssymb}
\usepackage{mathtools}
\usepackage{enumerate}
\usepackage{float}
\usepackage{bm}
\usepackage{upgreek}
\usepackage{rotating}
\usepackage{multicol}
\usepackage{array,multirow}
\usepackage{subfigure}
\usepackage{authblk}
\usepackage{textcomp}
\usepackage{graphicx}
\usepackage{caption}
\usepackage{algorithm}
\usepackage[noend]{algpseudocode}
\usepackage{url,hyperref}
\usepackage{amsthm}
\usepackage[left=2cm,right=2cm,top=2cm,bottom=2cm]{geometry}

\title{Deep learning and American options via free boundary framework}

\date{}
\numberwithin{equation}{section}

\author[1]{Chinonso Nwankwo \thanks{Corresponding author: \url{chinonso.nwankwo@ucalgary.ca
}}}
\author[2]{Nneka Umeorah}
\author[1]{Tony Ware}
\author[3]{Weizhong Dai}
\affil[1]{Department of Mathematics and Statistics, University of Calgary, Calgary T2N 1N4, Canada}
\affil[2]{School of Mathematics, Cardiff University, Cardiff CF24 4AG, United Kingdom}
\affil[3]{Department of Mathematics and Statistics, Louisiana Tech University, Ruston, LA 71272, USA}

\newtheorem*{remark}{Remark}
\begin{document}
\maketitle

\begin{abstract}
\noindent We propose a deep learning method for solving the American options model with a free boundary feature. To extract the free boundary known as the early exercise boundary from our proposed method, we introduce the Landau transformation. For efficient implementation of our proposed method, we further construct a dual solution framework consisting of a novel auxiliary function and free boundary equations. The auxiliary function is formulated to include the feed forward deep neural network (DNN) output and further mimic the far boundary behaviour, smooth pasting condition, and remaining boundary conditions due to the second-order space derivative and first-order time derivative. Because the early exercise boundary and its derivative are not a priori known, the boundary values mimicked by the auxiliary function are in approximate form. Concurrently, we then establish equations that approximate the early exercise boundary and its derivative directly from the DNN output based on some linear relationships at the left boundary. Furthermore, the option Greeks are obtained from the derivatives of this auxiliary function. We test our implementation with several examples and compare them with the existing numerical methods. All indicators show that our proposed deep learning method presents an efficient and alternative way of pricing options with early exercise features. \\

\end{abstract}

\section{Introduction}\label{sec1}
\noindent Different forms of neural networks (NN) have been developed to aid the solution of both low and high-dimensional PDEs. These include Physics-Informed NN \cite{Raissi, Raissii, Wang}; the feed-forward NN and its application in DeepXDE \cite{Lu}; convolutional NN \cite{Hex,Khoo}; recurrent NN \cite{Sabate}; Multi-layer perceptron (MLP) NN \cite{He}; radial basis function NN \cite{Golbabai, Jianyu}; modified artificial NN \cite{Hussian}; and the deep Galerkin Method for high-dimensional PDEs \cite{Sirignano}. Implementing the artificial NN for option pricing has been traced back to the works of Malliaris and Salchenberger  \cite{Malliaris} and Hutchinson et al.  \cite{Hutchinson}, whose research serves as a benchmark for the subsequent studies on the machine learning applications in pricing and hedging of financial derivatives. The authors showed that the European-style option valuation based on the Black-Scholes model could be successfully recovered with a neural network. DNN has been extensively studied in the area of pricing, hedging, and calibration of financial derivatives and stochastic models \cite{Chen}; implied volatility smile with its calibration, algorithmic trading, credit risk analysis \cite {Pacelli, Khashman}, assessment of bond ratings \cite{Dutta}, forecasting of volatility, etc. 
\\

\noindent Free boundary PDEs often appear in diverse science, engineering, and mathematics areas and have a wider application in heat transfer models, population models, predator-prey models, finance, fluid flow problems, etc. They can be seen in melting and solidification problem \cite{Alexiades, Ceseri, Hu, Kumar}, semi-conductor problem \cite{Morandi, Schmeiser}, forest fire simulation \cite{Filippi}, combustion processes \cite{Andreucci, Caffarelli, Frankel, Kim}, fluid-structure interactions \cite{Bansch}, logistic population models \cite{Piqueras}, and option pricing and hedging \cite{MacKean}. Most free boundary PDE models do not have a closed-form solution. Hence, several numerical and semi-analytical approaches have been proposed in the existing literature for solving these models. Recently, Wang and Perdikaris \cite{Wangg} proposed a deep learning approach for solving free boundary PDE related to the heat transfer problem. The authors presented a multi-network model based on physics-informed neural networks (PINN) and implemented it for solving one dimensional one-phase Stefan problem, one dimensional two-phase Stefan problem, and the two-dimensional one-phase Stefan problem. Also, Zhao et al. \cite{Zhao} solved a free boundary problem based on the Hele-Shaw model. The authors further proved and established the existence of their numerical solution.\\

\noindent American options model can be seen as a free boundary PDE model. In contrast to the vanilla European options, the American options possess the free boundary, and the holder of the options is allowed to exercise its right before expiration. This advantage comes with a cost because the early exercise boundary is computed simultaneously with the value function and the option Greeks. Some standard numerical and analytical approximation methods for solving the American options model include the binomial tree method \cite{Cox}, least square method \cite{Longstaff}, integral method \cite{Carr}, first passage approach \cite{Bunch, Gutierrez}, and mesh/meshless methods \cite{Brennan}. All these approaches are not without their advantages and limitations. In the PDE framework, American options can be formulated as a free boundary problem, in the linear complementary framework, or with a penalty term. The linear complementary framework and penalty method solve this model by introducing a constraint that removes the early exercise feature. It is important to mention that the early exercise boundary is an indispensable feature in American options valuation, which is very useful in the trading environment. Furthermore, the accuracy of the value function strongly depends on this early exercise feature. Computing the free boundary with high precision becomes paramount and not just a necessity. \\

\noindent The complication in solving American options becomes more substantial in a high dimensional context when precise early exercise boundary value is desired. For instance, American options with stochastic volatility involve a free boundary plane, a function of time and volatility. For American options with stochastic volatility and interest rate, a three-dimensional framework for the free boundary will be obtained:  a function of time, volatility, and interest rate. Hence, it might be cumbersome and computationally expensive to obtain the free boundary structure for the basket FX options with stochastic volatility when the finite difference, element, or volume method is considered. Furthermore, options that involve two or multi-free boundaries, like the strangle and straddle options, can also pose challenges to approximate in a high-dimensional context. When the conventional finite difference, volume, or element are implemented, considering multi-dimensional basket stock American options under a linear complementary framework can be computationally intractable due to the curse of dimensionality. It is worth mentioning that NN and its extension have also been implemented for solving American-style options in both low and high-dimensional contexts \cite{Anderson, Becker, Chen, Hirsa, Letourneau, Liu4, Reppen, Sirignano, Villani}.\\

\noindent Our research comes from the non-parametric aspect of option pricing and hedging as we seek to solve the American options model with deep learning approach. In this research, we propose a DNN method for solving the American options model as a free boundary problem which entails extracting the early exercise features while approximating the options simultaneously. To this end, we first use the logarithmic Landau transformation to fix the free boundary. Furthermore, we introduce a novel auxiliary function that both mimic the boundary values of the exact value function and in-cooperates the feed-forward DNN part. We formulate the auxiliary function to simultaneously, satisfies the far field boundary, the high contact condition known as the smooth pasting, and the second-order in space and first-order in time boundary conditions. These formulations are under the condition that a linear relationship between the early exercise boundary and the feed-forward DNN output at the left boundary holds. Training the neural network is an optimization procedure that involves computation and minimization of the gradient of the loss function obtained from the auxiliary function with respect to the network parameters. The performance of our proposed DNN approach is investigated and compared extensively with other numerical methods in the result section. Thus, the major highlights of this research are summarized below:
\begin{itemize}
\item We propose a novel auxiliary function and a hybrid implementation that solves the front-fixing American options model.
\item Due to the non-linearity and non-smoothness associated with the transformed American options model and number of our trainable data, we opt for the feed-forward DNN algorithm in approximating our value function, option Greeks, and early exercise boundary with its first derivative.
\item We perform some hyper-parameter tunings to determine the optimal architecture for our neural network model.
\item  Our implementation was tested with several examples that account for varying time to maturity $T$ and strike price $K$. We further compared our results with existing methods \cite{Egorova, Fazio, Nielsen} including the results obtained from a highly accurate sixth-order compact finite difference scheme with left boundary improvement \cite{Nwankwo}.
\end{itemize}

\noindent The remaining sections of this work are organized as follows. In Section \ref{sec2}, we introduced the transformed American options using the logarithmic Landau transformation and further normalized the transformed model. Section \ref{sec3} explains the concept of the DNN approximation. Section \ref{sec4} presents the experimental results; a discussion and analysis of the results follow, and then a comparison with the existing method is illustrated. Finally, Section \ref{sec5} concludes the study and provides guidelines for future research.  

\section{Normalized American Options Model with Landau Transformation} \label{sec2}
Let us consider a non-dividend-paying American put options written on an underlying asset. The asset price $S(t)$ is driven by geometric Brownian motion and given as
\begin{equation}\label{sde}
\mathrm{d}S(t) = \mu S(t) \mathrm{d}t + \sigma S(t) \mathrm{d}B(t).
\end{equation}
$\sigma$ and $\mu$ represent the volatility and the drift, respectively. If we denote $V(t,S)$ as the value function written on the asset price $S(t)$ with $K$, $T$, and $s_{f}(t)$ representing the strike price, time to maturity and the early exercise boundary, then the free boundary partial differential equation is then given as
\begin{equation}\label{f1}
\frac{\partial V(t,S)}{\partial t} - \frac{\sigma^2 S^2}{2} \frac{\partial^2 V(t,S)}{\partial S^2} -  S\frac{\partial V(t,S)}{\partial S}+rV(t,S)=0\,,\qquad S>s_{f}(t);
\end{equation}\
\begin{equation}\label{f2}
V(t,S))=K-S ,\qquad S<s_{f}(t).
\end{equation}
Here, we solve backwards in time and consider $t=T-t.$ The boundary and the initial conditions are given as follows:
\begin{equation}\label{f3}
V(0,S)= \max(K-S,0), \qquad V(t,s_{f}(t))=K-s_{f}(t)  \qquad V(t,\infty)=0.
\end{equation}
To locate and compute a more precise value of the early exercise feature with the options, we consider the logarithmic Landau transformation which was first introduced in the work of Wu and Kwok \cite {Wuu} for solving the American options model. To this end, let
\begin{equation}\label{f4}
 x=\ln S-\ln s_{f}(t), \qquad V(t,S)= U(t,x). 
\end{equation}
Hence, the transformed fixed free boundary problem with the initial and boundary condition is given as follows:
\begin{equation}\label{f5}
\frac{\partial U}{\partial t} - \frac{\sigma^2}{2} \frac{\partial^2 U}{\partial x^2} - \left(r+\frac{s'_{f}}{s_{f}}- \frac{\sigma^2}{2}\right)\frac{\partial U}{\partial x}+rU=0\,,\qquad x>0;
\end{equation}
\begin{equation}\label{f6}
U(t,x))=K-e^x s_{f}(t) ,\qquad x<0;
\end{equation}
\begin{equation}\label{f7}
 U(0,x)=0, \qquad U(t,0)=K-s_{f}(t),  \qquad U(t,\infty)=0,  \qquad t>0;
\end{equation}
\begin{equation}\label{f8}
 U_x(0,x)=0, \qquad U_x(t,0)=-s_{f}(t),  \qquad U_x(t,\infty)=0,  \qquad  t>0.
\end{equation}
Furthermore, some information regarding the second-order derivative of the option's value can be obtained from the partial differential equation when $x\rightarrow 0^{+}$ and $x\rightarrow \infty$ as follows:
\begin{equation}\label{f8a}
U_{xx}(t,0)= \frac{\ 2rK}{\sigma^2}-s_{f}(t),  \qquad U_{xx}(t,\infty)=0,  \qquad  t>0.
\end{equation}
After fixing the free boundary, we consider only the positive semi-infinite space domain. Hence the fixed space and time domain under consideration is given as $[0,\infty)\times [0,T]$. Because the option values vanish rapidly as we move further away from the fixed left boundary, if we consider the far field boundary condition with the right boundary given as $x_{max}$, then the fixed domain is reduced to $[0,x_{max}]\times [0, T]$. \\

\noindent For effective implementation of the DNN method for solving the transformed American options model, we reduced our fixed space and time domain from $[0,x_{max})\times [0,T]$ to $[0,1]\times [0,1]$. To this end, we introduce another transformation as follows:
\begin{equation}\label{f9}
 y = \frac{\ x}{x_{max}},  \qquad  \tau = \frac{\ t}{T}. 
\end{equation}
The normalized model with the fixed free boundary is then formulated as follows:
\begin{equation}\label{f10}
\frac{1}{T} \frac{\partial P}{\partial \tau} - \frac{1}{x_{max}^{2}}\frac{\sigma^2}{2} \frac{\partial^2 P}{\partial y^2} - \frac{1}{x_{max}} \left(r- \frac{\sigma^2}{2}\right) \frac{\partial P}{\partial y}- \frac{1}{x_{max}T}\frac{s'_{f}}{s_{f}}\frac{\partial P}{\partial y}+rP=0\,,\qquad y>0.
\end{equation}
For the initial and boundary conditions after the normalization of the transformed model,
\begin{equation}\label{f11}
P(\tau,y)=\frac{K}{x_{max}}-\frac{e^{yx_{max}}}{x_{max}} s_{f}(\tau) ,\qquad y<0;
\end{equation}
\begin{equation}\label{f12}
P(0,y)=0, \qquad P(\tau,0)=\frac{K}{x_{max}}-\frac{s_{f}(\tau)}{x_{max}}, \qquad P(\tau,1)=0, \qquad P(0,y\rightarrow 0^+)=0;
\end{equation}
\begin{equation}\label{f13}
 P_y(0,y)=0, \qquad P_y(\tau,0)=-s_{f}(\tau),  \qquad P_y(\tau,1)=0,  \qquad P_{y}(0,y\rightarrow 0^+)=0, \qquad  \tau>0.
\end{equation}
\begin{equation}\label{f13a}
P_{\tau}(\tau,0)=-\frac{s'_{f}(\tau)}{x_{max}},  \qquad P_{\tau}(\tau,1)=0,  \qquad  \tau>0.
\end{equation}
Substituting $P(\tau,y), P_y(\tau,y)$ and $P_{\tau}(\tau,y)$ into (\ref{f10}) when $y=0$, we then obtain the boundary value of the second-derivative of the value function as follows:
\begin{equation}\label{f14}
P_{yy}(\tau,0)= \left(\frac{\ 2rK}{\sigma^2}-s_{f}(\tau)\right)x_{max},  \qquad P_{yy}(\tau,1)=0, \qquad  \tau>0.
\end{equation}
Furthermore, for the non-dividend paying American put options pricing model, we have it that the early exercise boundary is monotonically decreasing with \cite{Musiela}
\begin{equation}
s_f(0) = K, \qquad 0\leq s_f(\tau)\leq K, \qquad s_f(\infty) = \frac{\xi}{\xi +1}K, \qquad \xi = \frac{2r}{\sigma^2}.
\end{equation}

\section{Front-Fixing DNN and the American Options Model}\label{sec3}
\noindent Deep learning (DL) is a sub-class of the machine learning algorithm \cite{Deng}, which uses multiple hidden layers to extract higher-level features from the specified raw input data. The learning can be supervised, unsupervised or semi-supervised and has applications in speech recognition, regression, image classification, game intelligence, etc. \cite{Liu2}. DL has attained remarkable success in different applied fields, though its application in PDEs has emerged recently. In this section, we present our proposed DNN method for solving the fixed free boundary American options, which simultaneously accounts for the early exercise feature, value function, and option Greeks.  

\subsection{DNN via an auxiliary function}
This section introduces a novel function that solves the optimal stopping problem and, more precisely, leverages the position to approximate the early exercise boundary simultaneously with the value function and Greeks using the DNN techniques. We called this function an auxiliary function. Closely related implementation was first introduced in the work of Lagaris et al. \cite{Lagaris} for solving ODEs and PDEs, which they called the trial solution. Subsequently, some authors implemented Lagaris et al. \cite{Lagaris} approach for solving the options pricing model mainly in the European options framework \cite{Eskiizmirliler, Hou, Umeorah}. Because we consider American options as a free boundary problem and implemented a front-fixing approach, several boundary values of the value function and its derivatives pose a more significant challenge using Lagaris et al. \cite{Lagaris} approach directly. This is by no means trivial, and we consider a slightly different approach that is similar in representation (but different in formulation) to the one presented in the work of Chen et al. \cite{Chenb} and Liu et al. \cite{Liu3}. To this end, we present the novel auxiliary function as follows:
\begin{align}\label{g1}
\mathcal{P}(\tau,y;\nu) &= e^{\frac{-y^2 \gamma}{2\tau}}\mathcal{M}(\tau,y;\nu).
\end{align}
The exponential part of $\mathcal{P}(\tau,y;\nu)$ closely resembles that of the probability density function of the Brownian motion. Here,
\begin{align}\label{gc1}
\mathcal{M}(\tau,y;\nu)&= \left[\frac{K-\bar{s}_f(\tau;\nu)}{x_{max}}-yK + \frac{y^2}{2}a(\tau;\nu) + yx_{\max}\mathcal{N}(\tau,y;\nu)  \right], \quad \gamma=\frac{r\phi}{\sigma^2},
\end{align}
\begin{align}\label{at}
a(\tau;\nu)&=\left(\frac{2rK}{\sigma^2}+ \frac{\gamma}{\tau x^2_{max}}(K-\bar{s}_f(\tau;\nu))-K+x_{max}\mathcal{N}(\tau,0;\nu)-2\mathcal{N}_y(\tau,0;\nu) \right)x_{max},
\end{align}
$\nu$ represents the adjustable weights, and biases from the DNN output $\mathcal{N}(\tau,y;\nu)$. Here, $\gamma$ controls the speed at which the auxiliary function vanishes when $y \rightarrow 1$, hence ensuring that we satisfy the far boundary behaviour. 

\begin{remark}
We select $\phi$ in such a way that $\gamma$ is not too small to force the exponential part to decay too slowly and also not too large which could result in extremely rapid decay. These two scenarios can hugely impact the far boundary behavior, minimization of the loss function, and the DNN approximation. We envisage that further analysis can be performed on $\phi$ to always determine the optimal value. With an extensive experiment, we observed that the range $90 \leq \phi \leq 110$ results in efficient minimization and provides a more stable DNN approximation. In the whole experiment we carried out in the result section, we use $\phi=100$.
\end{remark}

\noindent Furthermore, we uniquely formulate $\mathcal{P}(\tau,y;\nu)$ such that it simultaneously satisfy the payoff conditions, high contact condition known as the smooth pasting condition, and the remaining boundary values involving the first order time derivative and second order space derivative of the value function. We want to note that even though we formulate the auxiliary function to satisfy the exact boundary conditions, our left boundary values are not exact because of the early exercise boundary. This is because we simultaneously approximate the latter from the feed-forward DNN output with the value function. Here, the mathematical formulation in (\ref{g1}) holds under the condition that
\begin{align}\label{eb}
\bar{s}_f(\tau;\nu) &= K-x_{max}\mathcal{N}(\tau,0;\nu),\quad \bar{s}'_f(\tau;\nu) = -x_{max}\mathcal{N}_{\tau}(\tau,0;\nu).
\end{align}
The feed-forward DNN output and its derivatives are embedded both in the auxiliary function and free boundary equations. Our approach for approximating the early exercise boundary and its velocity is similar to the PINN method of Raissi et al. \cite{Raissii} when computing the boundary and initial values from the NN output. At this juncture, the inherent discontinuity in the option pricing model at the payoff is worth mentioning. To what extent the DNN output minimizes this challenge remains to be seen. From (\ref{f10}) and (\ref{g1}), one can easily observe that the transformed model and auxiliary function are dominated by the early exercise boundary and its first derivative. Furthermore, both the early exercise boundary and its derivative is approximated directly from the DNN output and its first derivative in time. Hence, the extent to which the DNN output learns and improves the inherent discontinuity in the model will influence the accuracy of the early exercise boundary and will further affect the precise computation of the option price and the Greeks. We give further details in the result section. It is easy to see from the auxiliary function that
\begin{align}\label{npcp1}
\mathcal{P}(\tau,0;\nu) &=\frac{K-\bar{s}_f(\tau;\nu)}{x_{max}}, \quad \mathcal{P}(\tau,y \rightarrow 1;\nu)\rightarrow 0, \quad \mathcal{P}(0,y;\nu)=0.   
\end{align}
Moreover, $\mathcal{P}(\tau,0;\nu)$ is the auxiliary function evaluated at $y=0$. $\mathcal{P}(\tau,y;\nu)$ will exactly have the same boundary values as $P(0,\tau)$ if $\bar{s}_f(\tau;\nu)=s_f(\tau)$. Taking derivatives of the auxiliary function, we then obtain
\begin{align}
\mathcal{P}_y(\tau,y;\nu) &= e^{\frac{-y^2 \gamma}{2\tau}}\left[-K + ya(\tau;\nu) + x_{max}\mathcal{N}(\tau,y;\nu) + yx_{max}\mathcal{N}_y(\tau,y;\nu)  \right] \nonumber
\\&\nonumber\\
& - \frac{y \gamma}{\tau} e^{\frac{-y^2 \gamma}{2\tau}}\left[\frac{K-\bar{s}_f(\tau;\nu)}{x_{max}}-yK + \frac{y^2}{2}a(\tau;\nu) + yx_{max}\mathcal{N}(\tau,y;\nu)  \right]. 
\label{g2}\\&\nonumber\\ 
\mathcal{P}_{\tau}(\tau,y;\nu) &= e^{\frac{-y^2 \gamma}{2\tau}}\left[\frac{-\bar{s}'_f(\tau;\nu)}{x_{max}} + \frac{y^2}{2}a'(\tau;\nu) + y*x_{max}\mathcal{N}_{\tau}(\tau,y;\nu)   \right]  \nonumber
\\&\nonumber\\
& + \frac{y^2\gamma}{2\tau^2} e^{\frac{-y^2 \gamma}{2\tau}}\left[\frac{K-s_f(\tau)}{x_{max}}-yK + \frac{y^2}{2}a(\tau;\nu) + yx_{max}\mathcal{N}(\tau,y;\nu)   \right]. \label{g3}
\end{align}
\begin{align}\label{conti}
a'(\tau;\nu)&=\left( -\frac{\gamma}{\tau^2 x^2_{max}} (K-\bar{s}_f(\tau;\nu)+\tau \bar{s}'_f(\tau;\nu))+x_{max}\mathcal{N}_{\tau}(\tau,0;\nu)- 2\mathcal{N}_{\tau y}(\tau,0;\nu) \right)x_{max}.
\end{align}
From the above equations, the following boundary conditions are satisfied:
\begin{align}\label{npcp2}
\mathcal{P}_y(\tau,0;\nu) &= -\bar{s}_f(\tau;\nu), \quad  \mathcal{P}_y(\tau,y \rightarrow 1;\nu)\rightarrow 0, \quad \mathcal{P}_y(0,y;\nu)=0, \quad \mathcal{P}_{\tau}(\tau,0;\nu) = \frac{-\bar{s}'_f(\tau;\nu)}{x_{max}}.
\end{align}
Finally, we differentiate the auxiliary function twice and obtain
\begin{align}\label{g4}
\mathcal{P}_{yy}(\tau,y;\nu) &= e^{\frac{-y^2 \gamma}{2\tau}} [a(\tau;\nu) + 2x_{max}\mathcal{N}_y(x,\tau;\nu)+yx_{max}\mathcal{N}_{yy}(\tau,y;\nu)]\nonumber
\\&\nonumber\\
& -2\frac{y \gamma}{\tau} e^{\frac{-y^2 \gamma}{2\tau}}\left[-K + ya(\tau;\nu) + x_{max}\mathcal{N}(\tau,y;\nu) + yx_{\max}\mathcal{N}_y(\tau,y;\nu)  \right] \nonumber
\\&\nonumber\\
& + \left(\frac{y^2\gamma^2}{\tau^2} -\frac{\gamma}{\tau}\right) e^{\frac{-x^2 \gamma}{2\tau}}\left[\frac{K-\bar{s}_f(\tau;\nu)}{x_{max}}-yK + \frac{y^2}{2}a(\tau;\nu) + yx_{\max}\mathcal{N}(\tau,y;\nu)  \right].
\end{align}
With L'Hôpital's rule and fixed $y$, one can observe that 
\begin{equation}
\lim_{\tau \to 0} \frac{y\gamma}{\tau}e^{-\frac{y^2 \gamma}{\tau}}=0, \quad \lim_{\tau \to 0} \frac{y^2\gamma}{\tau^2}e^{-\frac{y^2 \gamma}{\tau}}=0, \quad \lim_{\tau \to 0} \frac{y^2\gamma^2}{\tau^2}e^{-\frac{y^2 \gamma}{\tau}}=0,
\end{equation}
This is because the exponential part tends to zero faster than $\tau$. Furthermore, it is easy to see from \ref{g4} that the far boundary condition for the second derivative of the value function is satisfied. Moreover, we are now focused on the continuation region because we fix the free boundary. Hence, the discontinuity in the value function's first and second derivative (in space) is no longer visible; at most for $\tau>0$. We refer the reader to the work of Ballestra \cite{Ballestra}. To ensure that the second derivative of the auxiliary function satisfies the left boundary condition, we impose the boundary value of the second derivative of the value function on $\mathcal{P}_{yy}(\tau,0;\nu)$ as follows:
\begin{align}\label{g44}
\mathcal{P}_{yy}(\tau,0;\nu) &= \left(\frac{2rK}{\sigma^2}-\bar{s}_f(\tau;\nu) \right)x_{max},
\end{align}
and then obtain
\begin{align}\label{g5}
\left(\frac{2rK}{\sigma^2}-s_f(\tau) \right)x_{max} &= a(\tau;\nu)-\frac{\gamma}{\tau}\left[\frac{K-\bar{s}_f(\tau;\nu)}{x_{max}} \right]+2x_{max}\mathcal{N}_y(\tau,0;\nu).
\end{align} 
Thus, in order to satisfy (\ref{g5}), we define $a(\tau;\nu)$ as presented in (\ref{at}). If we observe the value function when $\tau = 0$, $K-s_f(0)=0$. Moreover, the velocity at which $s_f(0)\rightarrow K$ is extremely faster than the speed at which $\tau \rightarrow 0$. This attribute explains the main reason why the first derivative of the early exercise boundary is singular at payoff. Furthermore, the exponential part in $\mathcal{P}(\tau,y;\nu)$ and its derivative further ensure that the speed at which the auxiliary function tends to zero is faster than when $\tau \rightarrow 0$. We observed that we could achieve an accurate DNN solution if we infinitesimally avoid $\tau=0$. In this research, a stable solution is obtained up to $\tau_{min}=10^{-8}$. Finally, the auxiliary function can be represented as follows:
\begin{align}\label{g6}
\mathcal{P}(\tau,y;\nu)&= e^{\frac{-y^2 \gamma}{2\tau}}\left[\frac{K-\bar{s}_f(\tau;\nu)}{x_{max}}-yK + \frac{y^2}{2}\left( \frac{2rK}{\sigma^2}-\bar{s}_f(\tau;\nu) \right)x_{max} \right] \nonumber
\\&\nonumber\\ &  + \frac{y^2 \gamma  e^{\frac{-y^2 \gamma}{2\tau}}}{2\tau x_{max}} (K-\bar{s}_f(\tau;\nu)) +e^{\frac{-y^2 \gamma}{2\tau}}\left[yx_{max}\mathcal{N}(\tau,y;\nu) - y^{2}x_{max}\mathcal{N}_y(\tau,0;\nu) \right].
\end{align} 
With the formulation of (\ref{g6}), we have further ensured that the boundary values associated with the value functions and their derivatives present in(\ref{f10}) are fully mimicked in pseudo-exact form (due to the early exercise boundary) by the auxiliary function and its derivatives such that 
\begin{align}
\frac{1}{T} \frac{\partial \mathcal{P}(\tau,0;\nu)}{\partial \tau} - \frac{1}{x_{max}^{2}}\frac{\sigma^2}{2} \frac{\partial^2 \mathcal{P}(\tau,0;\nu)}{\partial y^2} - \frac{1}{x_{max}} &\left(r- \frac{\sigma^2}{2}\right) \frac{\partial \mathcal{P}(\tau,0;\nu)}{\partial y}\nonumber
\\&\nonumber\\
& -\frac{1}{x_{max}T}\frac{\bar{s}'_{f}}{\bar{s}_{f}}\frac{\partial \mathcal{P}(\tau,0;\nu)}{\partial y}+r\mathcal{P}(\tau,0;\nu)=0,
\end{align}
\begin{align}
\frac{1}{T} \frac{\partial \mathcal{P}(\tau,1;\nu)}{\partial \tau} - \frac{1}{x_{max}^{2}}\frac{\sigma^2}{2} \frac{\partial^2 \mathcal{P}(\tau,1;\nu)}{\partial y^2} - \frac{1}{x_{max}} &\left(r- \frac{\sigma^2}{2}\right) \frac{\partial \mathcal{P}(\tau,1;\nu)}{\partial y}\nonumber
\\&\nonumber\\
& -\frac{1}{x_{max}T}\frac{\bar{s}'_{f}}{\bar{s}_{f}}\frac{\partial \mathcal{P}(\tau,1;\nu)}{\partial y}+r\mathcal{P}(\tau,1;\nu) \cong 0,
\end{align}
\begin{align}
\frac{1}{T} \frac{\partial \mathcal{P}(\tau_{min},y;\nu)}{\partial \tau} - \frac{1}{x_{max}^{2}}\frac{\sigma^2}{2} \frac{\partial^2 \mathcal{P}(\tau_{min},y;\nu)}{\partial y^2} - \frac{1}{x_{max}} &\left(r- \frac{\sigma^2}{2}\right) \frac{\partial \mathcal{P}(\tau_{min},y;\nu)}{\partial y}\nonumber
\\&\nonumber\\
& -\frac{1}{x_{max}T}\frac{\bar{s}'_{f}}{\bar{s}_{f}}\frac{\partial \mathcal{P}(\tau_{min},y;\nu)}{\partial y}+r\mathcal{P}(\tau_{min},y;\nu)\cong 0.
\end{align}
Some of the Greek sensitivities are obtained directly from the auxiliary function and its derivatives after training. Furthermore, the early exercise boundary and its first derivative can easily be obtained from (\ref{eb}). The major difference between our auxiliary function and the trial solution of Lagaris et al. \cite{Lagaris} is that we formulate our function to simultaneously account for the fixed left boundary conditions and far boundary behaviours of the auxiliary function and its derivatives (up to second order derivative in space and first derivative in time. Furthermore, to account for the early exercise boundary and its derivative, we activate the Raissi et al. \cite{Raissi} PINN approach for dealing with the boundary values. Hence, our proposed deep learning method can be seen as a hybrid approach that introduces dual solution methods from the auxiliary function and the early exercise boundary (and its first derivative) equation for approximating the non-linear transformed American options model in (\ref{f10}).

\subsection{Grid sampling and refinements}\label{s33}
\begin{figure}[H]
    \centering
    \subfigure[Uniform grid sampling.]{\includegraphics[width=0.49\textwidth]{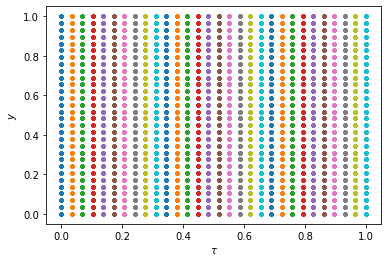}} 
    \subfigure[Randomly structured grid.]{\includegraphics[width=0.49\textwidth]{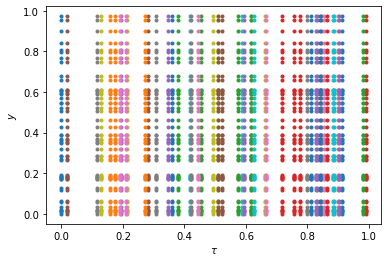}} 
    \subfigure[Grid stretching.]{\includegraphics[width=0.49\textwidth]{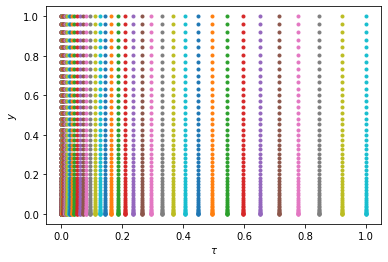}}    
    \caption{Grid sampling and refinement.}
    \label{grid}
\end{figure}

Solving the corresponding PDE on a discrete grid entails discretizing the domains and transforming the problem into an unconstrained minimization problem \cite{Lagaris}. Generally, uniform distribution and equispaced grid sampling techniques are popular in literature; others include Latin hypercube sampling, Sobol sequence, Halton sequence, and Hammersley sequence \cite{Wu}. This research considers uniform grid sampling, randomly structured grid, and grid stretching techniques for generating our training dataset (See Figure \ref{grid}) . It is important to note that the distribution of these training points greatly impacts our model's computational efficiency and accuracy, which in turn affects the minimization of the loss function, early exercise boundary, value function, and Greeks. \\

\noindent For the randomly structured grid, we first obtain a randomly spaced one-dimensional grid for the $y$ and $\tau$ input from the uniform distribution. We then use the NumPy \texttt{meshgrid} function to duplicate the grids and form a two-dimensional domain. Hence, this makes the randomly generated grid to be structured. The stretched grid sampling was generated from the uniform grid. In mesh methods like the finite difference, several methods have been proposed for improving the accuracy and convergence rate of the options pricing model. Some of these approaches use different grid refinement strategy that includes local mesh refinement, adaptive mesh refinement, and grid stretching based on coordinate transformation. The idea primarily concentrates the grid points close to the left boundary where singularity and discontinuity exist in the model. In this work, we implement a stretched grid sampling with a simple stretching as follows:
\begin{equation}
y_{\text{stretched}} = y^{\sqrt{p}}_{\text{uniform}},  \quad \tau_{\text{stretched}} = \tau^{p}_{\text{uniform}}, \quad p>1.
\end{equation}

\newpage
\subsection{Loss function and optimization}
\begin{figure}[H]
    \centering
{\includegraphics[width=0.7\textwidth]{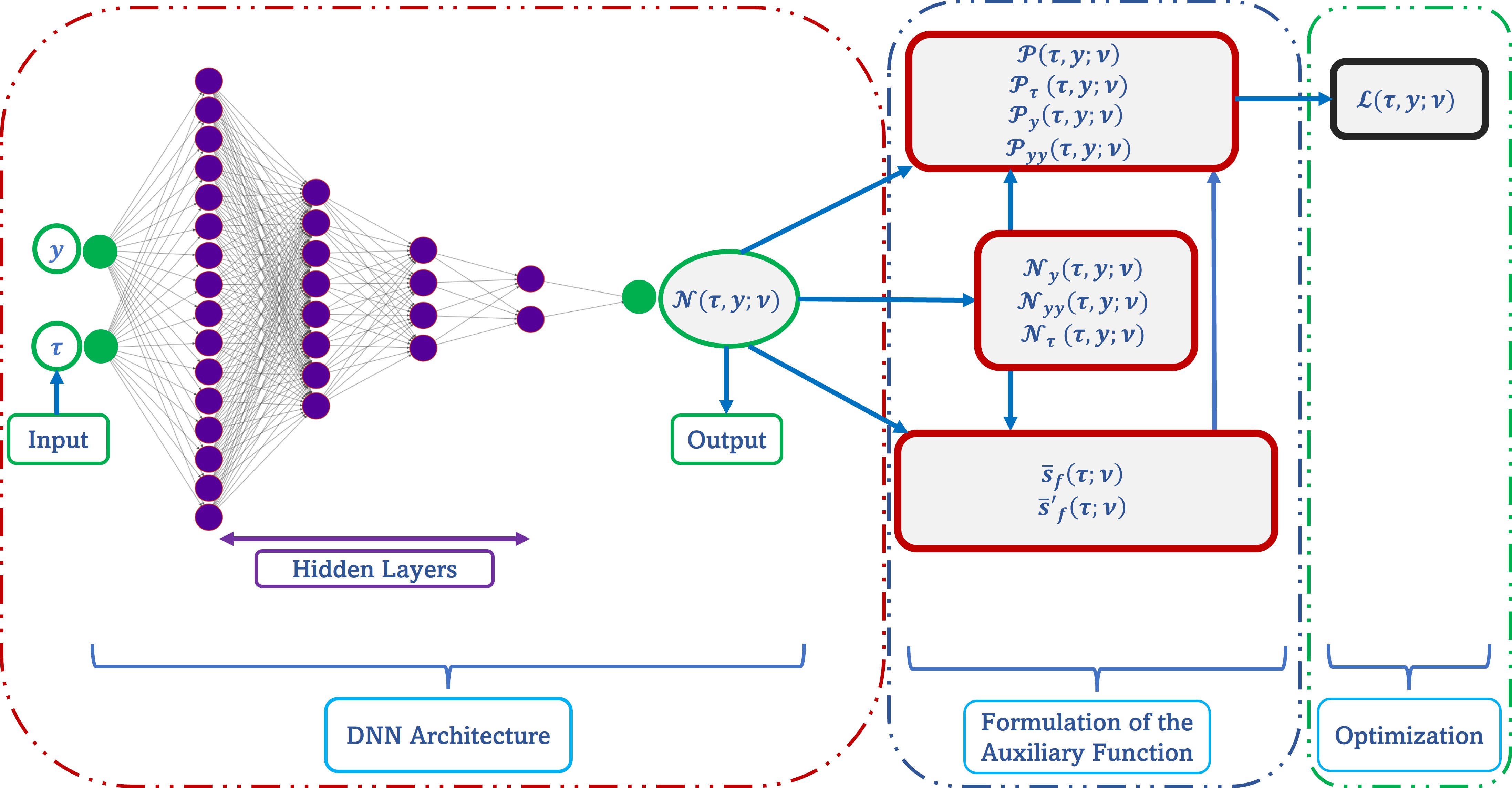}}  
    \caption{Flowchart of the DNN procedure.}
    \label{arch}
\end{figure}

\begin{figure}[H]
    \centering
    \subfigure[Less dense DNN architectures]
{\includegraphics[width=0.5\textwidth]{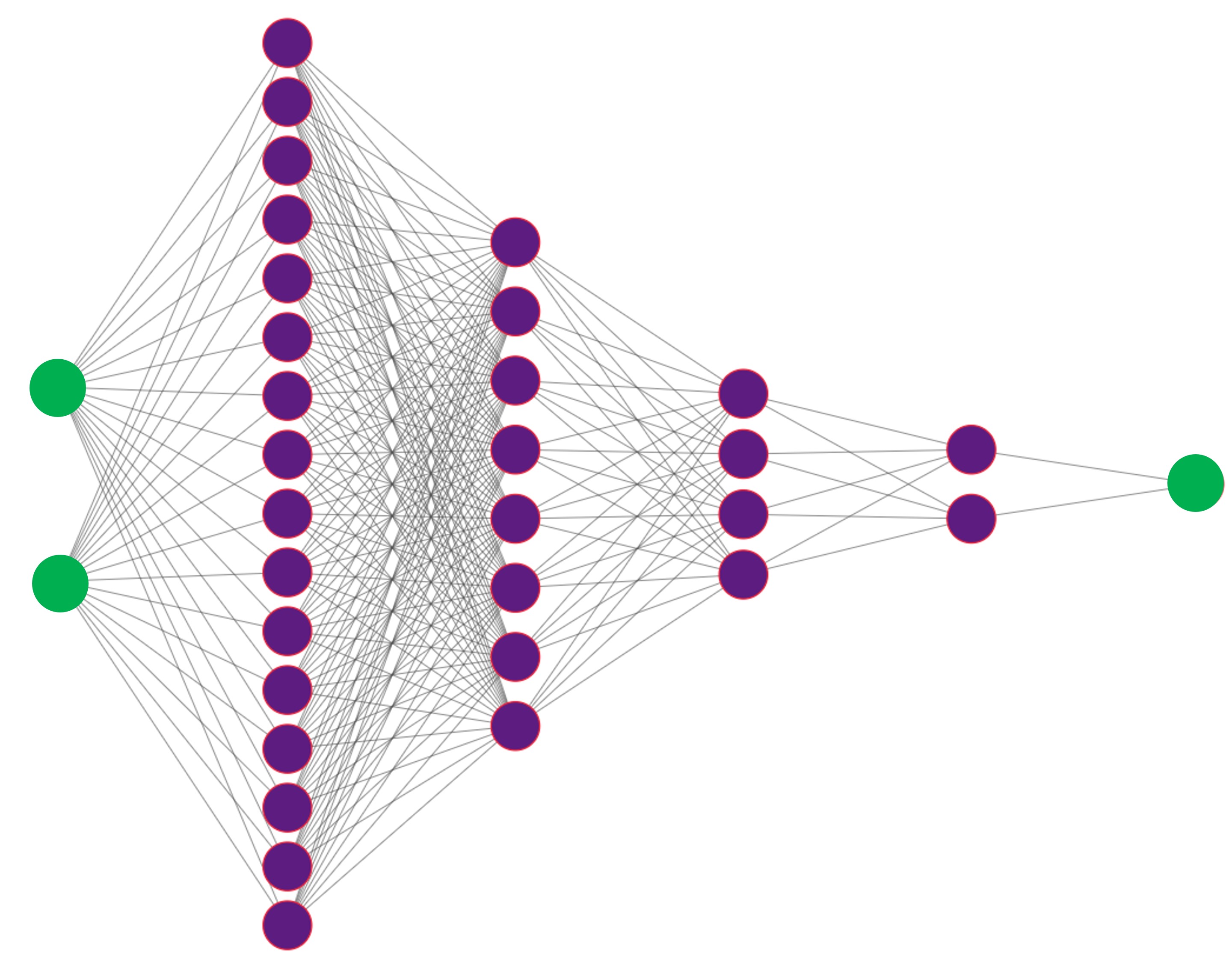}}  
    \subfigure[DNN architectures with uniform units]
{\includegraphics[width=0.5\textwidth]{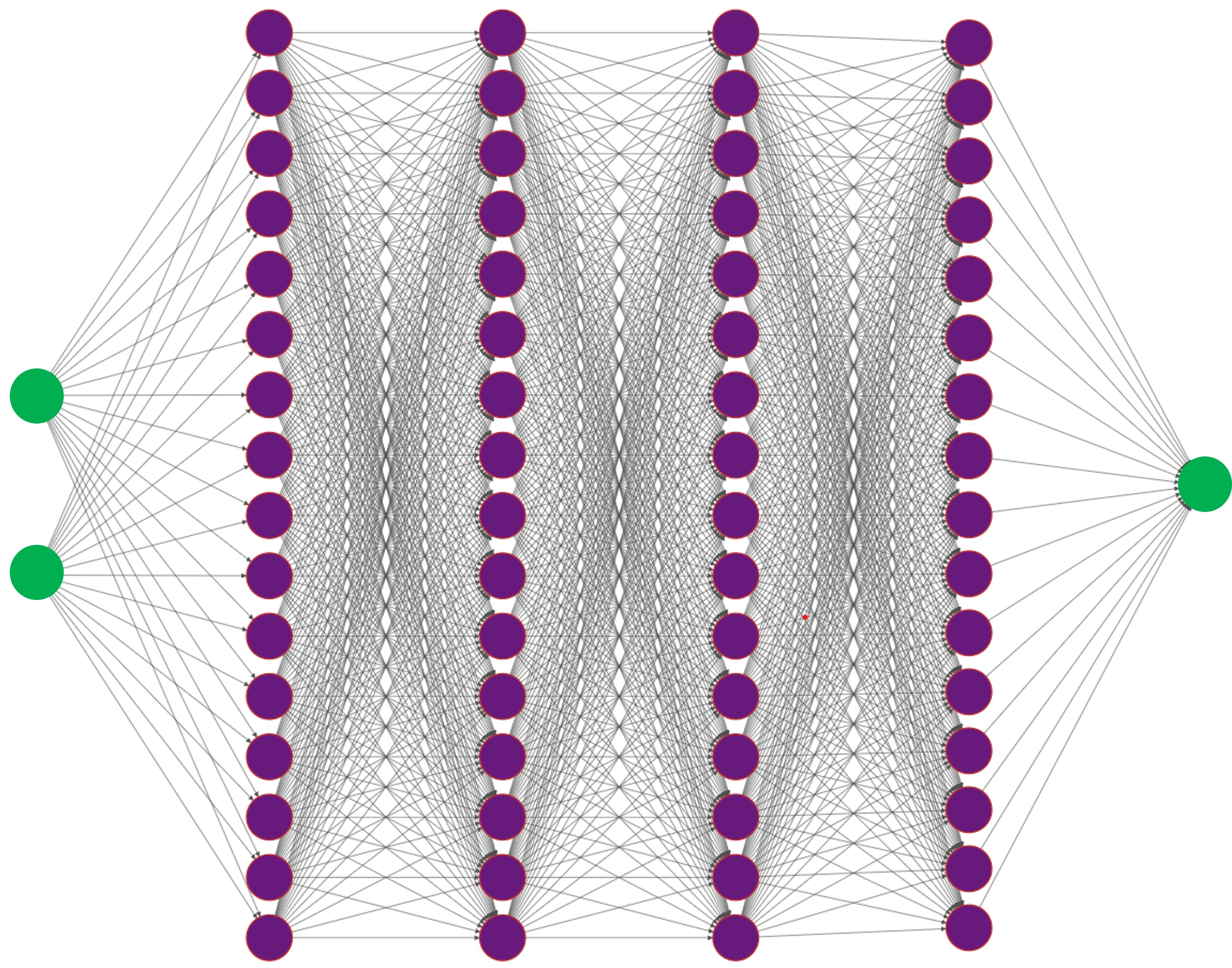}}
    \caption{Implemented DNN architectures \protect\footnotemark}
    \label{arch1}
\end{figure}

This section introduces the loss function and the optimization techniques employed in minimizing the loss function and training the DNN. We estimate the unknown adjustable DNN parameter $\nu$ by minimizing the following loss function:
\begin{equation}\label{lf}
\mathcal{L}(\tau,y;\nu)= \frac{1}{N_{\tau} N_{y}}\sum_{j=1}^{N_y}\sum_{i=1}^{N_{\tau}} \left[\frac{1}{T} \frac{\partial \mathcal{P}_{j,i}}{\partial t_i} - \frac{\sigma^2}{2x_{max}^{2}}\frac{\partial^2 \mathcal{P}_{j,i}}{\partial y_j^2} - \frac{1}{x_{max}} \left(r- \frac{\sigma^2}{2}+ \frac{1}{T} \frac{\bar{s}'_{f}}{\bar{s}_{f}}\right)\frac{\partial \mathcal{P}_{j,i}}{\partial y_j} +r\mathcal{P}_{j,i} \right]^2.
\end{equation}
for $i =0,1,\cdots, N_{\tau}$, $j=0,1,\cdots, N_y$ and $\mathcal{L}(\tau,y;\nu)$ is the error or cost function. The process can be constructed into an optimization problem
\begin{equation}\label{ke}
arg \min_{\nu} \mathcal{L}(\tau,y;\nu)\,.
\end{equation}
Training the DNN requires computing the gradient of the error function with respect to the input vectors and the DNN parameters. The partial derivatives of $\mathcal{P}(\tau,y;\nu)$ with respect to $y,\;yy$ and $\tau$ are already presented in equations (\ref{g2}), (\ref{g3}), and (\ref{g4}). Consider a feed-forward multi-layer perceptron with one hidden layer, then the linear output is given by
\begin{equation}\label{ou}
\mathcal{N}(\tau,y;\nu) = \sum_{i=1}^{M} \zeta_i f(w_i \tau + n_i y + B_i)\,,
\end{equation}
where $M$ is the total number of units in the DNN; $\zeta_i$ is the weight of the $i^{\text{th}}$ hidden unit; $w_i$ is the coefficient of the $\tau$ input to the $i^{\text{th}}$ hidden unit; $n_i$ is the coefficient of the $y$ input to the $i^{\text{th}}$ hidden unit; and $B_i$ is the bias value of the $i^{\text{th}}$ hidden neuron. Finally, for the hidden layers implemented in this work, we used $f(d)=(1+ \mathrm{e}^{-d})^{-1}$, the sigmoid activation function. The derivatives of $\mathcal{N}(y,\tau;\nu)$ with respect to $y,yy$ and $\tau$ are given below:
\begin{equation}
\mathcal{N}_{\tau}(\tau,y;\nu)=\sum_{i=1}^{M} \zeta_i w_i f(d_i) (1-f(d_i)),
\end{equation}
\begin{equation}
 \mathcal{N}_y(\tau,y;\nu)=\sum_{i=1}^{M} \zeta_i n_i f(d_i) (1-f(d_i)),
\end{equation}
\begin{equation}
\mathcal{N}_{yy}(\tau,y;\nu)=\sum_{i=1}^{M} \zeta_i n^2_i f(d_i) (1-f(d_i))(1-2f(d_i)).  
\end{equation}
Figures \ref{arch} and \ref{arch1} describes our deep learning procedure and less dense architecture, respectively. The following pseudocode (Algorithm~\ref{pseu}) explains the techniques for approximating the early exercise boundary and its first derivatives, value function and Greeks.

\begin{remark}
During the construction and the training of the DNN, weight initialization is essential to drive the convergence of the model efficiently. Failure to do this might lead to vanishing or exploding gradient issues. Also, since the NN methods are applied to a series of problems that are both highly non-linear and complex, then diverse initialization techniques tend to give rise to fluctuating performances. These techniques are presented in \cite{Dolezel}, and \cite{Narkhede}. \\

\noindent Weights are typically initialized to all zeros, all ones, from a normal distribution, uniform distribution, or even from techniques like the Xavier or Glorot Initialization. This research considered the random initialization from the normal distribution.
\end{remark}

\footnotetext{The authors credit the diagram for the DNN architecture in Figure \ref{arch1} to this link source: http://alexlenail.me/NN-SVG/index.html}

\begin{algorithm}[H]
\caption{Pseudocode for the optimal stopping problem implemented in \texttt{Python}}
\label{pseu}
Import the \texttt{tensorflow} library and other packages. Initialize the American options parameters.  
\begin{algorithmic}[1]

\State \textbf{Building the DNN Architecture}:
\begin{itemize}
\item  Initialize and discretize the training \& the prediction dataset into the required grid structure.
\item Initialize the parameters of the feed-forward DNN, such as the learning rate, number of the input, hidden, and output units.
\item Build the MLP and define the feed-forward flow $\mathcal{N}(\tau,y; \nu)$ with the prescribed architecture.
\end{itemize}

\State \textbf{Functions Description}: 

\begin{itemize}
\item Extract the early exercise boundary function from the DNN output, i.e. $\bar{s}_f(\tau;\nu) = K-x_{max}\mathcal{N}(\tau,0; \nu)$.
\item Extract the derivative of the early exercise boundary function from the first derivative (in time) of the DNN output, i.e. $\bar{s}'_f(\tau;\nu) = -x_{max}\mathcal{N}_{\tau}(\tau,0; \nu)$.
\item Construct the auxiliary function $\mathcal{P}(\tau,y; \nu) = e^{\frac{-x^2 \gamma}{2\tau}}\mathcal{M}(\tau,y; \nu)$ as shown in equation \ref{g1}; and construct the corresponding PDE as shown in equation \ref{f10} with their derivatives.
\end{itemize}  
\State \textbf{Training}:
\begin{itemize}
\item Define the Loss function $\mathcal{L}(\tau,y;\nu)$ in \ref{lf} with the exponentially decay learning rate; and the gradient $\Delta_{\nu}\mathcal{L}(\tau,y;\nu)$ of $\mathcal{L}(\tau,y;\nu)$ with respect to $\nu$.
\item Train the DNN and find optimal parameter $\nu'$: For the learning rate $q \in (0,1)$, adjust the parameters. \begin{align*}
\begin{cases}
              w \rightarrow w' &= w - q \mathcal{L}_w(\tau,y;\nu), \\
             B \rightarrow B' &= B - q \mathcal{L}_B(\tau,y;\nu). \,
            \end{cases}
\end{align*}
\end{itemize} 

\State  \textbf{Results}: Generate the test data points and compute the errors, value function, option Greeks, and the early exercise boundary.

\end{algorithmic}
\end{algorithm}

\section{Experiment and Discussion}\label{sec4}
The experiments presented in this work were performed on a computer with an 11th Gen Intel Core i7-1165G7 processor running at 2.80 GHz on a 64-bit Windows 10 operating system. Furthermore, we used the free and open-source software library of toolboxes 
 \texttt{Tensorflow (v 2.3.0)} \cite{Abadi} to construct the deep NN architecture and to train and optimize the model. To extensively verify the performance of our proposed DNN method for solving the fixed free boundary American option model, we first provide the result of our experiment, which we tested on a set of three different parameters consisting of short, medium, and long time to maturity $T$ and compare with the sixth order compact finite difference scheme. The data is provided in Table \ref{OptionVa} below.
\begin{table}[H]
\center
\caption{First experimental data.}
\label{OptionVa}
\begin{tabular}{|lccccc|}
 \hline 
Parameters & $K$ & $T$ & $r$  &$\sigma$ & $x_{max}$\\ 
 \hline \hline
Short Time to Maturity (STM) & 100    & 0.50  &0.05 & 0.20 & 6.00\\
Medium Time to Maturity (MTM) & 100    & 1.00 &0.10 & 0.30 & 6.00\\
Long Time to Maturity (LTM) & 100    & 3.00 &0.08 & 0.20 & 6.00 \\
\hline
 \end{tabular}  
\end{table}
\noindent Next, we will verify the performance of our proposed method with the parameter presented in the work of Nielsen et al. \cite{Nielsen} as displayed in Table \ref{OptionVaa}. The strike price $K$ in this second experiment is substantially smaller than the one considered in the first experiment as presented in Table \ref{OptionVa}. We compare the result obtained from Table \ref{OptionVaa} with ones obtained from the methods of Fazio et al. \cite{Fazio} (which involves repeated Richardson extrapolation), Nielsen et al. \cite{Nielsen} and Egorova et al. \cite{Egorova}.
\begin{table}[H]
\center
\caption{Second experimental data.}
\label{OptionVaa}
\begin{tabular}{|ccccc|}
 \hline 
 $K$ & $T$ & $r$  &$\sigma$ & $x_{max}$\\ 
 \hline \hline
 1.00    & 1.00  &0.10 & 0.20 & 2.00\\
\hline
 \end{tabular}  
\end{table}
\noindent Furthermore, it is worth mentioning that we slightly avoid $\tau=0$ when selecting our training dataset and approximation of our model. Here, we used $\tau_{min}=10^{-8}$. The reason for investigating our model's performance with several examples presented above is to verify and ensure that it can efficiently handle problems with a short, medium, and long time to maturity using varying parameters while validating its solution accuracy.

\subsection{Performance of grid samplings on the DNN solution}
In this subsection, we investigate the performance of the uniform, randomly structured, and stretched grid sampling on the solution accuracy of the proposed DNN method based on the above parameters. For the stretched grid, we verify its performance using several grid sets. In all the experiments involved in this work, we use a training step of 20000. \\

\begin{figure}[H]
    \centering
{\includegraphics[width=0.49\textwidth]{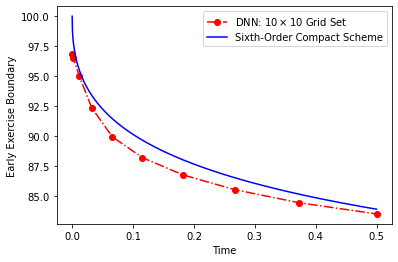}}
{\includegraphics[width=0.49\textwidth]{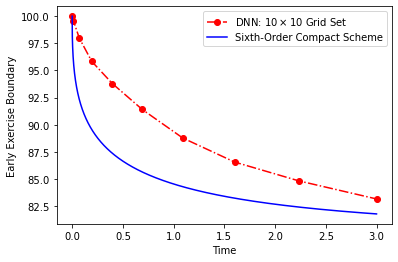}} 
{\includegraphics[width=0.49\textwidth]{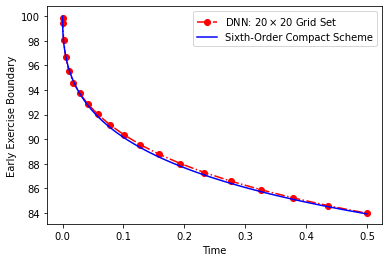}}  
{\includegraphics[width=0.49\textwidth]{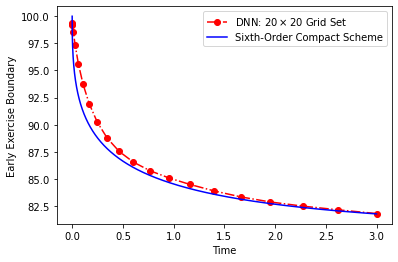}}  
{\includegraphics[width=0.49\textwidth]{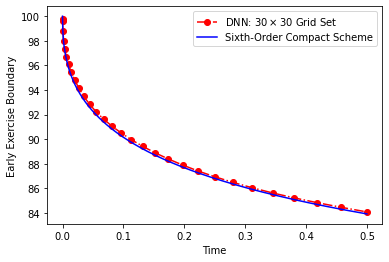}}
{\includegraphics[width=0.49\textwidth]{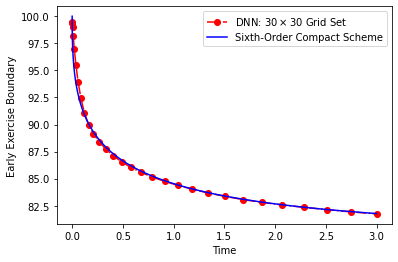}} 
{\includegraphics[width=0.49\textwidth]{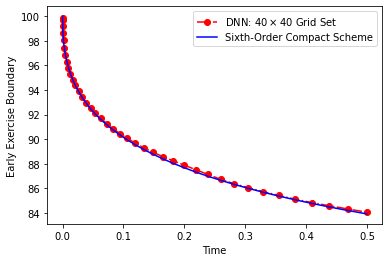}}  
{\includegraphics[width=0.49\textwidth]{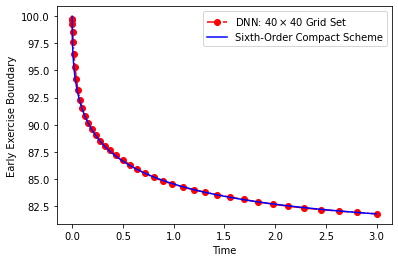}}
    \caption{Early exercise boundary in the LTM and STM scenarios using a stretched grid with $p=2.5$. STM is shown on the left side and LTM is in the right side. See Table \ref{OptionVa} for a description of the scenarios.}
    \label{fig:foobar2a}
\end{figure}

\noindent Furthermore, for the stretched grid, we ensure that $\tau_{min}^p = 10^{-8}$. Moreover, for the stretched grid solutions, across all subsection in this section, we use the stretched grid for training and uniform grid sampling for predicting. The only exception is in Figure \ref{fig:foobar2a} where we carried out both training and predicting procedure with the stretched grid to ensure we concentrate more predicting points near $\tau_{min}$. In this subsection, our DNN architecture consists of four hidden layers. The number of units decreases across the hidden layers progressively. We considered 1 input layer (2 × 1 neurons); 4 hidden layers (1 × 512 neurons, 1 × 256 neurons, 1 × 128 neurons, 1 × 64 neurons); and 1 output layer (1 × 1 neuron). The only exception is Figure \ref{fig:foobar2a} where we use 1 input layer (2 × 1 neurons); 4 hidden layers (1 × 256 neurons, 1 × 128 neurons, 1 × 64 neurons, 1 × 32 neurons); and 1 output layer (1 × 1 neuron). \\

\noindent Moreover, we observed that combining the Adam optimizer and sigmoid activation function with exponential learning rate decay minimizes our loss function more efficiently. For the Adam optimizer, the following parameters were used $\beta_1 = 0.99$, $\beta_2 = 0.999$, and $\epsilon = 10^{-8}$. It is important to mention, and as we will also see in the next subsection, that a slight change in $\beta_1$ greatly impacts the minimization of the loss function. The exponential learning rate decay ensures that the minimization of our loss function is consistent, especially after large training steps have been carried out. For the exponential learning rate decay, we used the decay step of 2000 and the decay rate of 0.85, which entails decaying every 2000 steps with a decay rate of 0.85.\\

\noindent In our first experiment, we verify the solution accuracy of the early exercise boundary and its first derivative obtained from our proposed DNN method when $\tau =T$ using the parameters in Table \ref{OptionVa}. The results are compared with the one obtained from the sixth-order compact finite difference scheme with third-order Runge-Kutta adaptive time stepping and free boundary improvement \cite{Nwankwo}. For the sixth order compact scheme, we use a step size of $h=0.005$. The results are display in Tables \ref{OptionVal2}-\ref{OptionVal4} and Figure \ref{fig:foobar2a}. From Tables \ref{OptionVal2}-\ref{OptionVal4}, the result using the stretched grid outperforms the ones obtained from the uniform grid and randomly structured grid sampling. For instance, the result obtained from the stretched grid with ($60\times60$) grid set is much closer to the benchmark result when compared with the uniform and structured grid sampling using the ($100\times100$) grid set. Even with a small training dataset, as displayed in Table \ref{OptionVal4}, the result from the stretched grid is still reasonably accurate compared to the benchmark value. We observed from our proposed DNN solution that we can achieve reasonable accuracy with a small training dataset if we concentrate more of those training points around the neighbourhood of the left boundary. Hence, the stretched grid sampling presents an optimal choice for our proposed DNN method. 

\noindent Furthermore, from Figure \ref{fig:foobar2a}, we observe that the plot of the early exercise boundary from the DNN output keep converging around the one obtained from the highly accurate sixth-order compact finite difference scheme with free boundary improvement as we increase the dataset. Furthermore, with grid stretching, a little increase in the training dataset is required to achieve reasonable convergence. It is interesting to see that with a stretched grid set of ($30\times30$), we have a reasonable match between the early exercise boundary obtain from our DNN approximation when compared with the one obtained from the sixth-order compact finite difference scheme. In our future work, we hope to investigate pure random grid sampling with low discrepancy as described in the work of Wu et al.\cite{Wu}. We hope to further investigate other grid refinement strategies that could improve the overall solution accuracy of our DNN approximation.
\begin{table}[H]
\center
\caption{Early exercise boundary with (100$\times$100) grid set ($\tau = T$).}
\label{OptionVal2}
\begin{tabular}{|ccccc|}
 \hline 
Method & Uniform & Random & Stretched ($p=2.0$) & Sixth-order compact scheme \\ 
 \hline \hline
STM &  84.27 & 84.52  &84.02 & 83.92 \\
MTM & 76.19    & 76.23  &76.16 & 76.16 \\
LTM & 81.84    & 81.78  &81.78 & 81.78 \\
\hline
 \end{tabular}  
\end{table}

\begin{table}[H]
\center
\caption{Early exercise boundary with different stretched (60$\times$60) grid set ($\tau = T$).}
\label{OptionVal3}
\begin{tabular}{|cccc|}
 \hline 
 \textbf{$p$} & 1.2 & 1.5 & 1.7 \\ 
 \hline \hline
 STM & 84.20    & 84.04  & 83.91 \\
 MTM & 76.26    & 76.17  &76.16  \\
 LTM & 81.88    & 81.80  &81.78  \\
\hline
 \end{tabular}  
\end{table}

\begin{table}[H]
\center
\caption{Early exercise boundary with stretched grid ($p=2.5,\;\tau = T$).}
\label{OptionVal4}
\begin{tabular}{|cccc|}
 \hline 
 Grid set & $(20\times20)$ & $(30\times30)$ & $(40\times40)$  \\ 
 \hline \hline
 STM & 84.22    & 84.06  &84.03  \\
 MTM & 76.19    & 76.18  &76.17  \\
 LTM & 81.76    & 81.78  &81.77  \\
\hline
 \end{tabular}  
\end{table}
\noindent In the second experiment, we present the DNN result we obtained with the parameter in Table \ref{OptionVaa} which we compared with the existing methods. The result is displayed in Table \ref{OptionVal2a}. We observed that our result with a stretched grid sampling ($p=2.5$) is the same as the ones in the works of Fazio et al. \cite{Fazio} and Egorova et al. \cite{Egorova} up to 5 digits and Nielsen et al. \cite{Nielsen} up to 4 digits. It is also quite interesting that we achieved this result with a very small training data set i.e., (20$\times$20) grid set. Hence, we further validate the importance of the stretched grid sampling on the solution accuracy of our DNN approximation.

\begin{table}[H]
\center
\caption{Early exercise boundary with (20$\times$20) grid set.}
\label{OptionVal2a}
\begin{tabular}{|ccccc|}
 \hline 
Methods & Nielsen et al \cite{Nielsen} & Fazio et al. \cite{Fazio} & Egorova et al. \cite{Egorova} & DNN solution ($p=2.5$)\\ 
 \hline \hline
$s_f(T)$ &  0.8622  & 0.8628  &0.8628 & 0.8628 \\
\hline
 \end{tabular}  
\end{table}
 
\subsection{Hyperparameter tuning, exponential learning rate decay and the loss function}\label{hyp}

Hyperparameter tuning is an indispensable step in the ML algorithm, as this is essential for the ML algorithm to output the result with the highest accuracy. There are diverse ways of tuning the hyperparameters, including manual search, grid search, Bayesian optimization, etc. We employ the manual search optimization technique for this research, despite being intensive and exhaustive. Here, we check the performance of the Adam optimizer\footnote{Adam is an optimization algorithm based on adaptive estimation of momentums of the first order and second order\cite{Kingma}. It has the reputation of converging rapidly and being computationally efficient.} on the loss function minimization for our DNN model. We slightly tune the values of the Adam optimizer and investigate how it impacts on the loss function and solution accuracy. The first set of approaches we employed in this research was to examine the impact of using/not using the exponential learning rate decay on the training optimizer, and this was performed on the same number of the hidden layer, neurons, and across the ($60 \times 60$) grid set. (For the architecture, see Table \ref{OptionVal5}).

\begin{table}[H]
\center
\caption{Hyperparameter tuning with stretched grid and ($60\times60$) grid set.}
\label{OptionVal5}
\begin{tabular}{|lccc|}
 \hline 
Hyperparameter  & Method 1 & Method 2 & Method 3\\ 
 \hline \hline
Hidden Layer & 4  & 4  & 4 \\
Units & (512,256,128,64)  & (512,256,128,64) & (512,256,128,64)  \\
Learning rate & 0.001  & 0.001 & 0.001  \\
Training step & 20000  & 20000 & 20000  \\
Display step & 1000  & 1000  & 1000  \\
Decay step & 2000  & none  & 2000 \\
Decay rate & 0.850  & none  & 0.850 \\
$\beta_1$ & 0.990  & 0.900  & 0.900  \\
$\beta_2$ & 0.999  & 0.999 & 0.999 \\
$\epsilon_{adam}$ & $10^{-8}$  & $10^{-8}$  & $10^{-8}$  \\
\hline
 \end{tabular}  
\end{table}
\begin{figure}[H]
    \centering
{\includegraphics[width=0.49\textwidth]{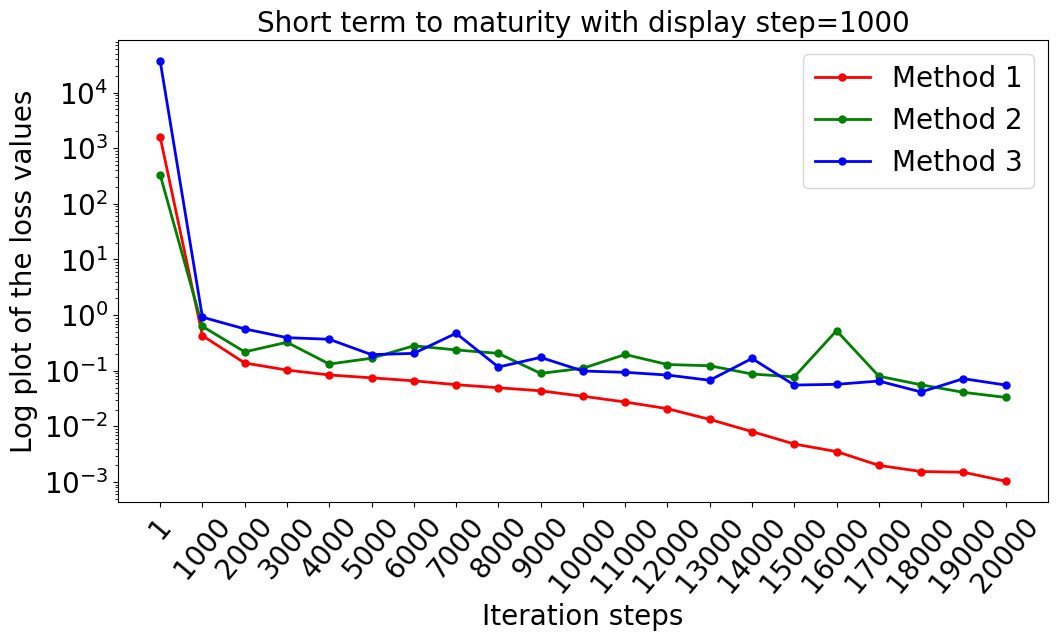}} 
{\includegraphics[width=0.49\textwidth]{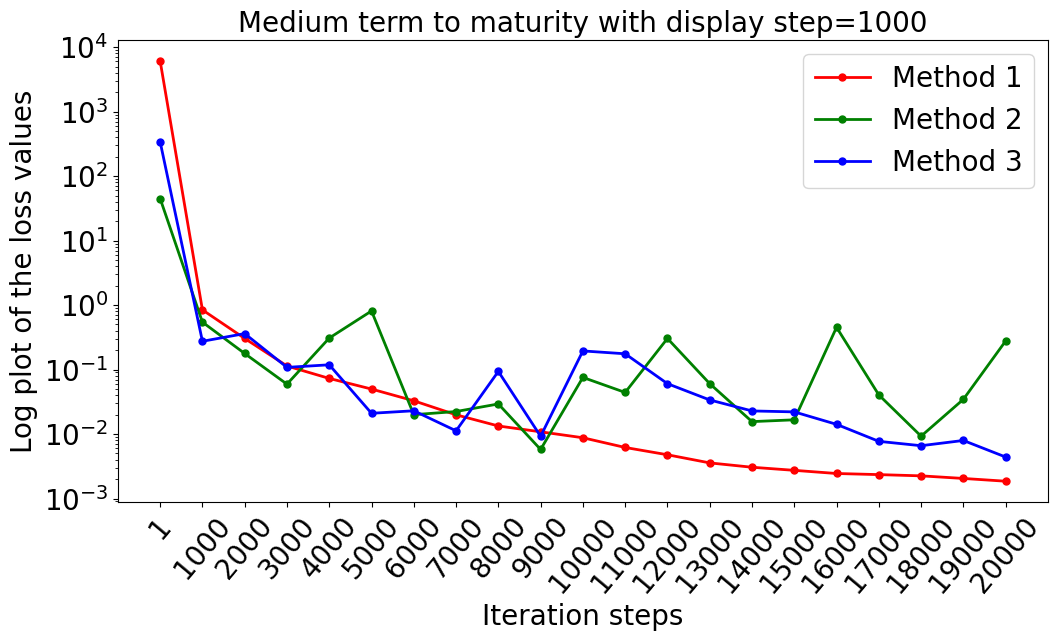}} 
{\includegraphics[width=0.49\textwidth]{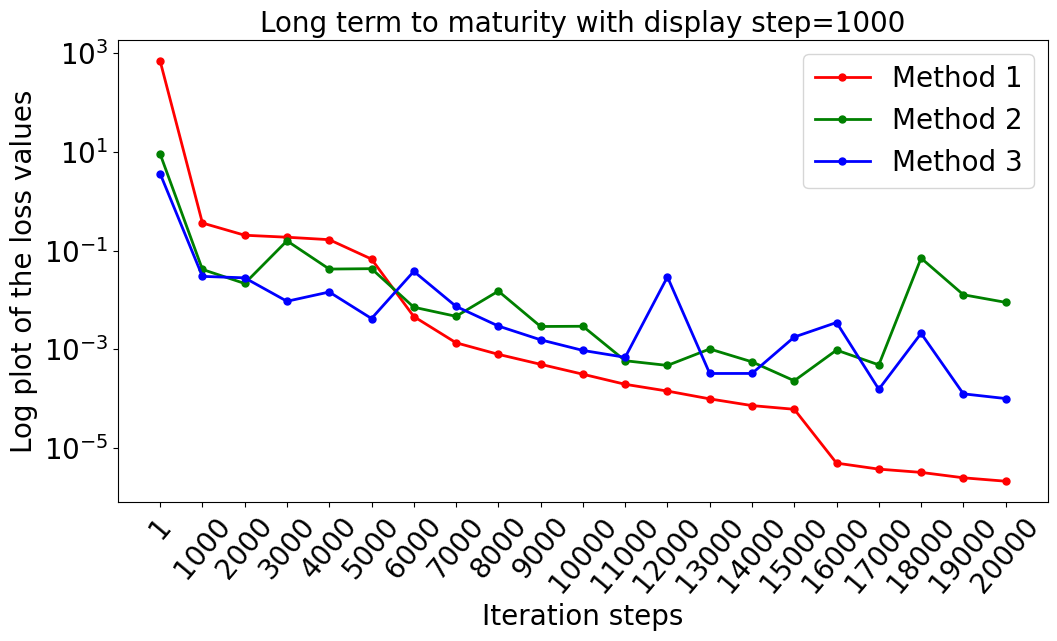}}    
    \caption{Log plot of loss function values for various time to maturity.}
    \label{fig:foobar}
\end{figure}

\noindent We visualized these results across all time to maturity, i.e., the STM, MTM, and the LTM. The result we obtained from the comparison showed that the exponential rate decay significantly impacted the loss function and the solution accuracy. For Methods 2 and 3, the oscillations were quite intense, and this behaviour affected the final minimized values and the accuracy of the solution. Thus, it was observed that the exponential learning rate decay trained the network with the specified significant learning rate, then slowly decayed the rate until the realization of the local minima. Furthermore, it can be seen that the conventional Adam parameter might not be optimal for all models, implying that it could be model dependent. Hence, Method 1 performed better when compared to Methods 2 and 3, as displayed in Figure \ref{fig:foobar}. The substantial improvement observed in the plot profile of Method 1 based on the implementation of the exponential learning rate and Adam parameter tuning was also reported in the work of Anderson and Ulrych \cite{Anderson}.\\

\begin{figure}[H]
    \centering
    \subfigure[Final loss values for different NN methods.]{\includegraphics[width=0.49\textwidth]{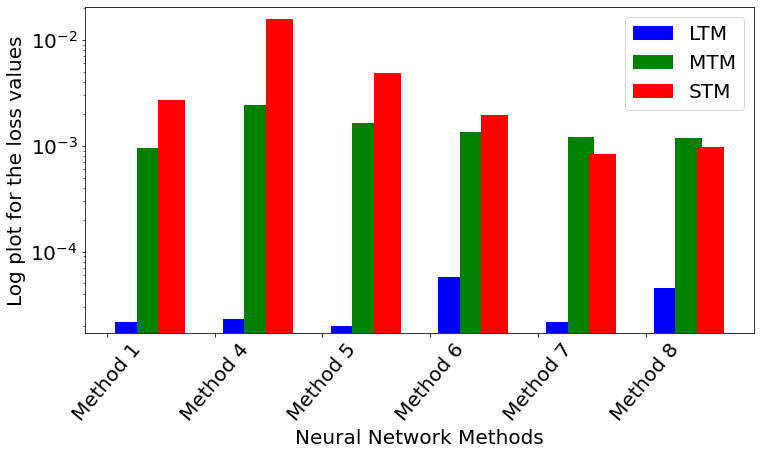}} 
    \subfigure[Computational time for different NN methods.]{\includegraphics[width=0.49\textwidth]{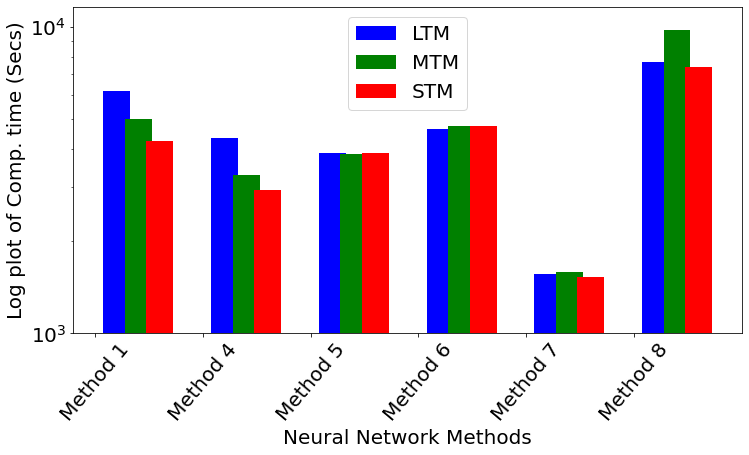}} 
    \caption{Comparison of loss function values and computational time with the last training step.}
    \label{fig:foobars}
\end{figure}

\noindent Next, we investigate the impact of the loss function when we implement Method 1 and then compare it with the results of implementing Methods 4-8. For Table \ref{OptionVal6}, we vary and increase the hidden layers and the units for Methods 4, 5, and 6. We equally fix the hidden layers and progressively decrease the number of units across the hidden layers for Methods 7 and 8, as displayed in Table \ref{OptionVal7}. After the training of the NN, we choose the last loss function values for Method 1 and Methods 4-8 above. We compare the performance based on the lowest loss function values and the computational time of each of the algorithms and display these results in Figure \ref{fig:foobars}. Regarding the plot for the loss function values (STM) in Figure \ref{fig:foobars}, the value for Method 4 was significantly larger compared to the other methods, whereas Method 7 gave the least loss value. Thus, there is a percentage decrease of 94.78\% when Method 4 (the highest) was compared to Method 7 (the lowest). Furthermore, we measured the impact of loss values on the MTM; we observed that the loss values were relatively small across all methods, though Method 1 gave the least of all. Similarly, for the LTM, the loss values were significantly small across all methods, though Method 5 gave the least of all. 

\begin{table}[H]
\center
\caption{DNN architecture with fixed units and varying hidden layers.}
\label{OptionVal6}
\begin{tabular}{|lccc|}
 \hline 
Hyperparameter  & Method 4 & Method 5 & Method 6 \\ 
 \hline \hline
Hidden Layer & 4  & 5 & 6  \\
Units & (200,200,200,200)  & (200,200,200,200,200) & (200,200,200,200,200,200)  \\
Learning rate & 0.001  & 0.001 & 0.001  \\
Training step & 20000  & 20000 & 20000   \\
Display step & 1000  & 1000  & 1000  \\
Decay step & 2000  & 2000 & 2000   \\
Decay rate & 0.850  & 0.850 & 0.850    \\
$\beta_1$ & 0.990  & 0.990  & 0.990 \\
$\beta_2$ & 0.999  & 0.999 & 0.999 \\
$\epsilon_{adam}$ & $10^{-8}$  & $10^{-8}$  & $10^{-8}$   \\
\hline
 \end{tabular}  
\end{table}
\begin{table}[H]
\center
\caption{DNN architecture with fixed hidden layer and varying units.}
\label{OptionVal7}
\begin{tabular}{|lcc|}
 \hline 
Hyperparameter  & Method 7 & Method 8 \\ 
 \hline \hline
Hidden Layer & 4  & 4   \\
Units & (256,128,64,32)  & (768,384,192,96)   \\
Learning rate & 0.001  & 0.001   \\
Training step & 20000  & 20000   \\
Display step & 1000  & 1000   \\
Decay step & 2000  & 2000   \\
Decay rate & 0.850  & 0.850   \\
$\beta_1$ & 0.990  & 0.990   \\
$\beta_2$ & 0.999  & 0.999  \\
$\epsilon_{adam}$ & $10^{-8}$  & $10^{-8}$    \\
\hline
 \end{tabular}  
\end{table}
\noindent For the computational time, irrespective of the time to maturity, Method 7 was computationally inexpensive compared to other methods, and this was a result of fewer hidden layers and units used in building the model. Consider the MTM as an example; the following percentage decrease was observed when all the other methods were compared to Method 7: 68.4\% (for Method 1), 51.02\% (for Method 4), 58.9\% (for Method 5), 66.8\% (for Method 6) and 83.9\% (for Method 8). Furthermore, we observe the impact of using fixed or varying units across the hidden layers. It is easy to observe the effect of this implementation based on the result obtained from Methods 4 and 7 in Figure \ref{fig:foobars}. We observed that when we use a fixed unit of (200,200,200,200) based on Method 4, the computational time is very high compared with Method 7 with varying units of (256,128,64,32). However, both of them gave almost the same final loss function value. When Method 4 was further compared to Method 7, the following percentage decrease was observed: 94.78\%, 51.02\%, and 6.55\% for STM, MTM, and LTM, respectively \footnote {It is essential to mention that all the plot profiles in section \ref{hyp} are in log scale. It enables better visualization of the LTM profile for the final loss function value and the computational time.}. Hence with ($60 \times 60$) grid set, we chose Method 7 as optimal in the DNN architecture. This choice was due to the less computational time and low loss function value we achieved, and this should be considered while avoiding overfitting and model complexity.

\subsection{Impact of far field boundary and near payoff conditions}
It is well known that some errors are introduced when a far-field boundary is considered in the American options model. Moreover, by introducing  $\tau_{min}$; a point very close to the terminal point, another source of error could be introduced further. Here, with fixed $\tau_{min} = 10^{-8}$, we first vary the value of $x_{xmax}$ and observe its impact on the solution accuracy of the early exercise boundary. Next, we vary the value of $\tau_{min}$ and carry out the same observation. For this experiment, we use Method 7, which gave us the shortest computational time and last loss function value in the previous subsection. We further compare the result with the one obtained with Method 1 and the sixth-order compact finite difference scheme as shown in Tables \ref{OptionVal2} and \ref{OptionVal3}. The result is displayed in Table \ref{impact}.
\begin{table}[H]
\center
\caption{Impact of $x_{max}$ and $\tau_{min}$ on the early exercise boundary with ($60\times 60$) grid set ($p=1.7$).}
\label{impact}
\begin{tabular}{|ccccc|}
 \hline 
Maturity time& $x_{max}=6$& $x_{max}=4$ & $x_{max}=3$ & $x_{max}=2$ \\ 
\hline \hline
STM & 83.91   &84.04 & 83.81 &83.95  \\
MTM & 76.17  &76.17  &76.19 &79.92   \\
LTM &81.80 &81.79  &83.19  &90.39  \\
\hline\hline
& $\tau_{min}=10^{-8}$& $\tau_{min}=10^{-6}$ & $\tau_{min}=10^{-4}$ & $\tau_{min}=10^{-2}$ \\
\hline \hline
STM & 83.91 &83.86   &83.88   &83.89  \\
MTM & 76.17 &76.21  &76.20  & 76.22  \\
LTM & 81.80 &81.78  &81.76 &81.83    \\
\hline
 \end{tabular}  
\end{table}

\noindent When we compare the result in Table \ref{impact} with the one obtained from the sixth order compact finite difference, it is easy to see that with fixed $\tau_{min}=10^{-8}$, we have reasonable accuracy when $x_{max}\geq 4$. Furthermore, if we fix $x_{max}= 6$, we observed that $\tau_{min}=10^{-8}$ gives the optimal accuracy with STM, MTM, and LTM. We observed that even though Method 1 gives a more accurate result than  Method 7 when compared with the sixth-order compact finite difference, the difference in the result is not substantial. Moreover, the computational time and last loss function value obtained from Method 7 is very substantial compared to Method 1, as shown in Figure \ref{OptionVal5}. Hence, we can validate that with a $60\times 60$ grid set, Methods 1 and 7 present an optimal choice of parameter required to obtain a reasonably accurate solution of the early exercise boundary with less computational time when compared with Methods 2-6, 8, and the sixth-order compact finite difference scheme. 

\subsection{Impact of the auxiliary function and DNN output on the boundary values}
Here, we verify the importance of our proposed novel auxiliary function which satisfies in pseudo-exact form, the boundary conditions imposed in the transformed and normalized American options. Our focus is on the left boundary point. Two common approaches exist in the literature for accounting for the boundary and initial condition in the loss function. One accounts for the boundary in the exact form and the other is the PINN approach where the boundary and initial values are generated directly from the NN output. Chen et al. \cite{Chenb} mentioned that though the latter is easier to implement and extend to the high-dimensional problem and complicated domain, this approach does not always lead to exact satisfaction of the boundary and initial conditions and might require more computational effort. Our formulation as presented in (\ref{g1}) and (\ref{gc1}) is a novel hybrid approach that integrates both approaches mentioned above for the efficient and simultaneous approximation of the early exercise boundary, value function, and Greeks. Simply, we use the first approach to present boundary values that are in pseudo-exact form due to the early exercise boundary and then use the PINN approach to approximate the early exercise boundary and its first derivative directly from the feed-forward DNN output.\\

\begin{figure}[H]
    \centering
    \subfigure[Case 1 (\ref{at1}).]{\includegraphics[width=0.48\textwidth]{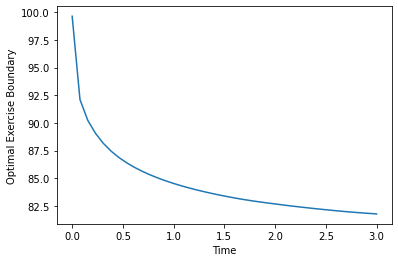}} 
    \subfigure[Case 2 (\ref{conti1}).]{\includegraphics[width=0.48\textwidth]{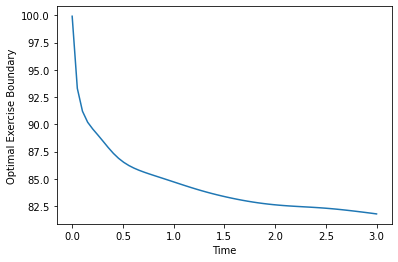}} 
    \caption{Plot of the early exercise boundary from two different boundary scenarios.}
    \label{fig:foobarss}
\end{figure}

\noindent In this subsection, we present an alternative implementation where we assume we do not know the boundary value of the second derivative of the value function when $y=0$ and extract the information directly from the DNN output by differentiating the latter twice with respect to $y$. Moreover, for this implementation, we take cognizant of the fact that the Greek variables are inherently sensitive which will be more pronounced at the left boundary. To this end, we then present two different boundary scenarios and compare their performance extensively. Case 1 represents our formulation in (\ref{g1})-(\ref{conti}), where we satisfy the boundary values in the pseudo-exact form as follows:
 
\begin{align}\label{at1}
\mathcal{P}_y(\tau,0;\nu) = -\bar{s}_f(\tau;\nu), \quad \mathcal{P}_{yy}(\tau,0;\nu) &= \left(\frac{2rK}{\sigma^2}-\bar{s}_f(\tau;\nu) \right)x_{max}.
\end{align}
In Case 2, the left boundary value of the second derivative of our auxiliary function is directly obtained from the feed-forward DNN output as follows:
\begin{align}\label{conti1}
\mathcal{P}_y(\tau,0;\nu) = -\bar{s}_f(\tau;\nu), \quad \mathcal{P}_{yy}(\tau,0;\nu) &= \mathcal{N}_{yy}(\tau,0;\nu),
\end{align}
with
\begin{equation}
a(\tau;\nu)=\left(\frac{\gamma}{\tau x_{max}}(K-\bar{s}_f(\tau;\nu))+\mathcal{N}_{yy}(\tau,0;\nu)-2x_{max}\mathcal{N}_y(\tau,0;\nu) \right).
\end{equation}
\begin{align}
a'(\tau;\nu)&=\left( -\frac{\gamma}{\tau x_{max}} (K-\bar{s}_f(\tau;\nu)+\tau \bar{s}'_f(\tau;\nu))+\mathcal{N}_{\tau yy}(\tau,0;\nu)- 2x_{max}\mathcal{N}_{\tau y}(\tau,0;\nu) \right).
\end{align}
We compare the performance of these two boundary scenarios on the solution accuracy and the final loss function value using LTM, Method 1, ($60\times 60$) stretched grid set with $p=2.5$ and 20000 training steps. The results are displayed in Table \ref{perform} and Figure \ref{fig:foobarss}.

\begin{table}[H]
\center
\caption{Solution accuracy and final loss function value based on (\ref{at1}) and (\ref{conti1})}
\label{perform}
\begin{tabular}{|ccc|}
 \hline 
 Parameter & Case 1 (\ref{at1}) & Case 2 (\ref{conti1}) \\ 
 \hline \hline
 $s_f(T)$ & 81.773    & 81.788 \\
 Final loss function value & $4.437\times 10^{-3}$     & $4.170\times 10^{-2}$  \\
\hline
 \end{tabular}  
\end{table}

\noindent From Table (\ref{perform}), we observe that the final loss function value is lower in Case 1 when compared with Case 2. Chen et al. \cite{Chenb} further mentioned in their literature that due to the sensitivity of the boundary and initial conditions, our approach in Case 1 can reduce the effort required during the training of the neural network. Moreover, as shown in Figure \ref{fig:foobarss} the early exercise boundary plot profile obtains from Case 1 in is much smoother when we compare it to Case 2. Furthermore, the value of the early exercise boundary obtain using the sixth-order compact finite difference scheme is 81.777. We observed that the early exercise boundary obtained from Case 1 in (\ref{at1}) is more accurate when compared with the one obtained from Case 2 in (\ref{conti1}). Hence, we validate that our novel auxiliary function in (\ref{g1}) provides a more accurate solution. \\

\noindent However, the importance of Case 2 cannot be undermined especially when the boundary value of the second derivative is not known or accounted for. For example, in the existing literature that solves the American-style stochastic volatility model with the front-fixing approach, the boundary value for the second derivative was not accounted for \cite{Zhu}. The approach implemented in Case 2 can be used to generate the boundary value of the second derivative from the feed-forward DNN output and the latter can be compared with any existing exact boundary value under investigation. This phenomenon is left for future research and we hope to study it further when solving free boundary options involving stochastic volatility and interest rate using our proposed deep learning method.

\subsection{Value function and Greek sensitivities}
\begin{figure}[H]
    \centering
    \subfigure[Value function in three-dimensional axis.]{\includegraphics[width=0.48\textwidth]{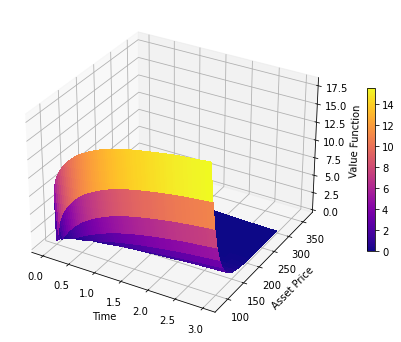}} 
    \subfigure[Value function with varying $\tau$.]    
{\includegraphics[width=0.48\textwidth]{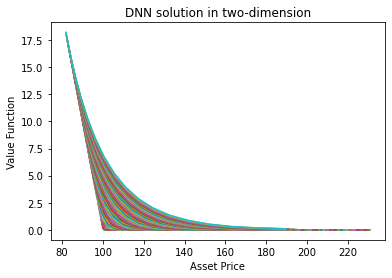}} 
    \subfigure[Value function when $\tau=0$ and $\tau = T$.]    
{\includegraphics[width=0.48\textwidth]{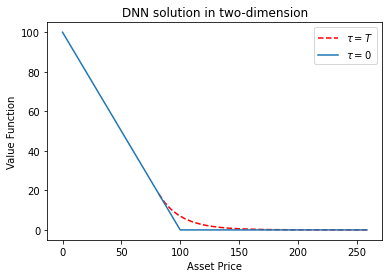}}
    \subfigure[Comparison of the value function.]   
{\includegraphics[width=0.48\textwidth]{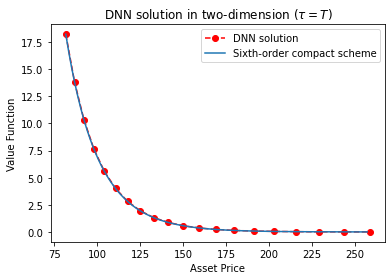}}    
    \caption{Solution of the value function with LTM.}            
    \label{fig:foobar3}
\end{figure}

\begin{figure}[H]
    \centering
    \subfigure[Delta sensitivity in three-dimensional axis.]   
{\includegraphics[width=0.48\textwidth]{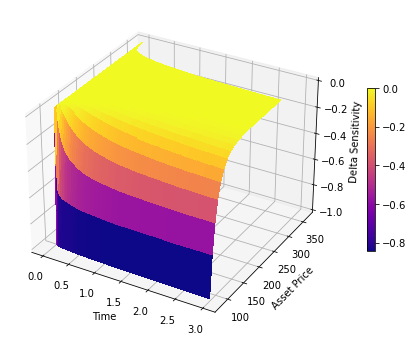}} 
    \subfigure[Comparison of the Delta sensitivity.]    
{\includegraphics[width=0.48\textwidth]{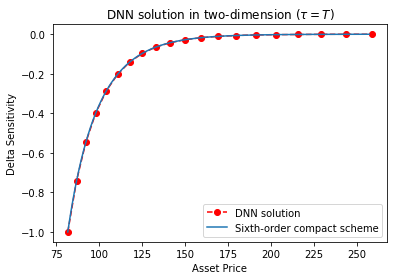}}
    \caption{Solution of the Delta sensitivity with LTM.}            
    \label{fig:foobar4}
\end{figure}

\begin{figure}[H]
    \centering
    \subfigure[Gamma sensitivity in three-dimensional axis.]   
{\includegraphics[width=0.48\textwidth]{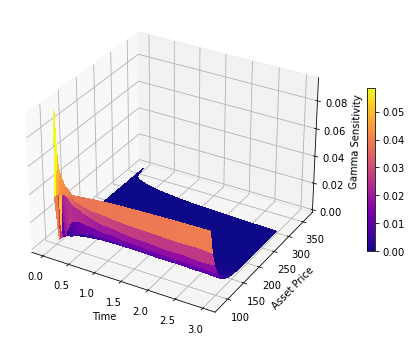}}    
    \subfigure[Gamma sensitivity in two-dimensional axis.]   
{\includegraphics[width=0.48\textwidth]{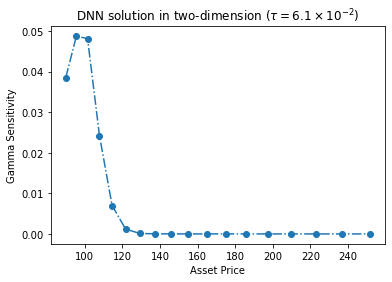}}
    \caption{Solution of the Gamma sensitivity with LTM.}            
    \label{fig:foobar5}
\end{figure}

In this last subsection, we verify the solution accuracy of our proposed DNN method with respect to the option value and Greek sensitivities. We further compare our DNN result with the one obtained from the sixth-order compact finite difference scheme.  The value function and option Greeks are computed from auxiliary function and its derivatives based on (\ref{g1}-\ref{g4}). The plot profile of the value function and the option Greeks are displayed in Figures  \ref{fig:foobar3}-\ref{fig:foobar5}.\\

\noindent From Figures \ref{fig:foobar3} and \ref{fig:foobar4}, the plot of the value function and Delta sensitivity from the DNN solution coincides reasonably well with the one from the sixth-order compact finite difference scheme. Similarly, we observed the same for the delta sensitivity. We did not compare the result of the Gamma sensitivity, however, the smoothness of the latter can easily be observed from Figure \ref{fig:foobar5}. In conjunction with the result we obtain from other subsections, we can validate that our proposed DNN method could provide a more efficient approach for solving the free boundary option pricing PDE model which could be very useful in a high dimensional context.
\vspace{-1mm}
\vspace*{1mm}

\section{Conclusion}\label{sec5}
We have proposed a deep learning method for solving the American options model as a free boundary problem. For the efficient computation of the early exercise boundary and its derivative simultaneously with the value function and the Greeks, we first implement a logarithmic Landau transformation. We then introduce a novel auxiliary function that includes the feed-forward DNN part and mimic the boundary behaviors. We trained this auxiliary function and its derivative to simultaneously account for the smooth pasting condition, far boundary behavior, and other boundary conditions associated with the derivatives of the value function under the further condition that a linear relationship between the early exercise boundary and feed-forward DNN output at the left boundary holds. \\

\noindent The performance of our DNN method was rigorously tested with several examples and compared with the existing methods. All indicators show that our proposed DNN method provides a more efficient and alternative way of solving free boundary style options and could be very useful in a high-dimensional context. Furthermore, our approach can easily be extended for solving other free boundary PDEs like the Stefan problem \cite{Liu5}, logistic population model \cite{Piqueras}, etc., where Landau transformation can be used for extracting the free boundary features.\\

\noindent Even though our experiment is almost exhaustive, in our future work, we will further enhance the performance of our proposed method by implementing a more rigorous hyperparameter tuning approach that will give us an optimal DNN architecture such that we can obtain more precise solution and efficient minimization with lesser computational time. Also, we will conduct a convergence analysis of our deep learning method for solving our model and carry out an extensive sensitivity analysis to extract the remaining Greeks. We also hope to extend our approach to American options with jumps, regime switching, and stochastic volatility and interest rate, energy-related options with one or multi-free boundaries, free boundary Quanto and basket FX options, and other exotic options that involve more than one free boundary like straddle and strangle options.

\section*{Acknowledgement}
The first author is funded in part by an NSERC Discovery Grant. The second author was funded in part by PIMS and the School of Mathematics, Cardiff University during her research visit to the University of Calgary.

%

\end{document}